\documentclass{iopjournal}

\usepackage[numbers, sort&compress]{natbib}
\usepackage{amsmath}
\usepackage{amssymb}
\usepackage[utf8]{inputenc}
\usepackage[T1]{fontenc}
\usepackage{booktabs}
\usepackage{dcolumn}
\usepackage{subcaption}
\usepackage{tabularx}
\usepackage{lmodern}
\usepackage{CJKutf8}

\newcolumntype{d}[1]{D{.}{.}{#1}}

\newcommand{\cF}{\mathcal{F}}

\begin{document}

\articletype{Paper}

\title{Mitigation of Incoherent Spectral Lines via Adaptive Coherence Analysis for Continuous Gravitational-Wave Searches}

\author{%
  Ye Zhou (\begin{CJK*}{UTF8}{gbsn}周烨\end{CJK*})$^{1,\dagger}$\orcid{0009-0004-7050-7736} and
  Karl Wette$^{1,2,*}$\orcid{0000-0002-4394-7179}
}

\affil{$^1$Centre for Gravitational Astrophysics, Australian National University, Canberra ACT 2601, Australia}

\affil{$^2$Australian Research Council Centre of Excellence for Gravitational Wave Discovery (OzGrav), Hawthorn VIC 3122, Australia}

\affil{$^\dagger$E-mail: ye.zhou.horizon@gmail.com}

\affil{$^*$E-mail: karl.wette@anu.edu.au}


\begin{abstract}
  The sensitivity of continuous gravitational-wave searches is strictly limited by non-Gaussian spectral artefacts that accumulate coherent power over long observation baselines. In this paper, we present an unsupervised mitigation framework based on adaptive network coherence analysis. Unlike traditional veto methods that discard entire frequency bands, our pipeline selectively suppresses local artefacts while preserving global potentially astrophysical signals. We validate the method using Advanced LIGO O3 data, analysing the cleaning performance across subsets spanning 1, 3, and 5 days. For the subsets spanning five days of O3 data, the pipeline identifies and mitigates 89\% and 77\% of the total spectral lines in the Hanford and Livingston detectors, respectively, while effectively preserving the coherent population consistent with astrophysical morphologies. This is achieved while modifying less than 7\% of the analysis bandwidth spanning 20~Hz to 2000~Hz. Rigorous statistical verification and recovery of simulated signals demonstrate that the mitigation effectively suppresses the non-Gaussian tail of the noise distribution while strictly preserving the statistical integrity of coherent signal candidates. By recovering detector sensitivity in parameter spaces previously contaminated by the spectral forest, this framework provides a robust preprocessing strategy for all-sky searches.
\end{abstract}

\section{Introduction}

The completion of the fourth observing run (O4) of the Advanced LIGO~\cite{theligoscientificcollaborationAdvancedLIGO2015}, Advanced Virgo~\cite{acerneseAdvancedVirgoSecondgeneration2015}, and KAGRA~\cite{akutsuOverviewKAGRADetector2021} detectors marks a major milestone in gravitational-wave (GW) astronomy. With the detectors having become more sensitive during this run~\cite{soniLIGODetectorCharacterization2025}, the global network is now undergoing critical upgrades~\cite{observing-plans} and paving the way for next-generation facilities like Cosmic Explorer~\cite{evansHorizonStudyCosmic2021} and the Einstein Telescope~\cite{maggioreScienceCaseEinstein2020}. Since the first GW signal from the merger of two black holes was detected in 2015~\cite{abbottObservationGravitationalWaves2016}, the catalogue of transient compact binary coalescences (CBCs) continues to grow, now comprising hundreds of black hole and neutron star mergers~\cite{abbottGW170817ObservationGravitational2017, abbottGWTC1GravitationalWaveTransient2019, abbottPopulationMergingCompact2023, abbottGWTC3CompactBinary2023, abbottGWTC21DeepExtended2024, collaborationObservationGravitationalWaves2024}.

The observation of continuous gravitational waves (CWs) remains, however, a key missing discovery. As highlighted in recent reviews~\cite{rilesRecentSearchesContinuous2017, SienBejg2019-CntGrvWNtSCrSPr, TenoEtAl2021-SMtCnGrvSgUSAdvE, Picc2022-StPrsCnGrvWSr, wetteSearchesContinuousGravitational2023, HaskBejg2023-AstCnGrvWv, JoneRile2025-MltObsSECnWTPrEqSDM, owenColloquiumMultimessengerAstronomy2026}, detecting these persistent signals would be scientifically invaluable. The most likely source of CWs are rapidly-rotating neutron stars. A CW detection would provide a direct probe into the interior dynamics of such stars, complementing observations of binary neutron star mergers. They would place unique constraints on the equation of state~\cite{laskyGravitationalWavesNeutron2015, glampedakisGravitationalWavesSingle2018}, and reveal details about crustal deformations~\cite{ushomirskyDeformationsAccretingNeutron2000} and r-mode instabilities~\cite{anderssonNewClassUnstable1998} that are inaccessible to electromagnetic astronomy~\cite{jonesGravitationalWavesRotating2002, bildstenGravitationalRadiationRotation1998}. To find these weak signals, researchers have employed a wide range of algorithms. Coherent matched-filtering searches commonly use the $\cF$-statistic~\cite{jaranowskiDataAnalysisGravitationalwave1998, CutlSchu2005-GnrFsMlDtMGrvWP, Prix2007-SrCnGrvWMMltFs, wetteFlatParameterspaceMetric2013}, which quantifies the consistency of the data with a specified CW signal model after analytically maximising over the unknown signal-amplitude parameters. Larger values of the $\cF$-statistic indicate greater consistency with the assumed CW signal model. Other approaches include semi-coherent methods such as the Hough transform~\cite{krishnanHoughTransformSearch2004, astoneMethodAllskySearches2014} and PowerFlux~\cite{Derg2011-DscPw2AlImp, Derg2012-LsChrSrSWllSg}, as well as Hidden Markov Model (HMM) tracking for signals with wandering frequencies~\cite{suvorovaHiddenMarkovModel2016, sunHiddenMarkovModel2018, jonesSearchContinuousGravitational2021}.
Despite applying these complex pipelines to the extensive Advanced LIGO and Virgo datasets, for example in all-sky surveys of the O3 (third)~\cite{abbottAllskySearchContinuous2022} and O4~\cite{collaborationAllskySearchContinuous2025} observing runs, no CW detection has been confirmed to date.

CW emission is modelled as a simple sinusoid in the reference frame of the neutron star. The frequency of the sinusoid slowly decays over time, as GWs extract energy from the rotational energy of the star. When observed at Earth, the signal acquired a further Doppler modulation due to the relative motion between the neutron star and the detector, chiefly due to the sidereal and orbital motion of the Earth. The changing directional sensitivity of the GW as it rotates with the Earth introduces further amplitude modulation of the signal. Notwithstanding these effects, CW signals are typically characterised as quasi-sinusoidal signals, and would therefore appear as narrow features in a Fourier transform of the GW data.

The sensitivity of CW searches is very often impeded by the non-Gaussian nature of GW detector data. While the fundamental noise floor is governed by seismic gradients, suspension thermal noise, and quantum vacuum fluctuations~\cite{buikemaSensitivityPerformanceAdvanced2020}, the data is also extensively populated by narrow-band spectral artefacts~\cite{davisLIGODetectorCharacterization2021, acerneseVirgoO3Run2022, abbottGuideLIGOVirgo2020}. Unlike transient CBC signals, CW searches integrate data over months to years, a regime where even low-amplitude, stationary instrumental or environmental lines accumulate significant coherent power and thereby masquerade as a quasi-sinusoidal CW signal~\cite{covasIdentificationMitigationNarrow2018}. This spectral forest contains both globally coherent features and a vast population of local, incoherent disturbances. Prominent examples include the 60~Hz U.S. mains power harmonics~\cite{nguyenEnvironmentalNoiseAdvanced2021}, the ``violin'' modes of the silica suspensions~\cite{cummingDesignDevelopmentAdvanced2012}, and calibration lines, including photon calibrator (Pcal) actuation lines, which couple with calibration control loop resonances~\cite{karkiAdvancedLIGOPhoton2016, vietsReconstructingCalibratedStrain2018, sunCharacterizationSystematicError2020}. Furthermore, ``fast scatter'' noise from stray light coupling continues to degrade data quality in the most sensitive bands~\cite{soniReducingScatteredLight2021, soniModelingReductionHigh2024}. These artefacts pose a severe challenge to established pipelines. For coherent methods based on the $\cF$-statistic, stationary noise lines frequently intersect with the Doppler-modulated frequency trajectory of a target source, generating false candidates that are computationally expensive to veto~\cite{keitelSearchContinuousGravitational2014, zhuNewVetoContinuous2017, JoneEtAl2022-VlCnGrvCnSmcSUDMEPSF}. Similarly, in semi-coherent schemes such as HMM tracking, wandering instrumental lines often resemble the stochastic spin-down of a neutron star, necessitating frequency masking that sacrifices valuable parameter space~\cite{suvorovaHiddenMarkovModel2016, sunHiddenMarkovModel2018}.

Recent progress has been made to apply deep learning to CW searches, ranging from early convolutional neural networks~\cite{dreissigackerDeeplearningContinuousGravitational2019} to recent transformer architectures~\cite{joshiTransformerNetworksContinuous2025} that achieve essentially matched-filter sensitivity in controlled studies. Deploying these models in complex detector environments, however, remains challenging. As demonstrated in the recent open data-analysis competition~\cite{tenorioLearningDetectContinuous2025}, semi-coherent matched filtering remains the dominant strategy for maximising sensitivity. The competition revealed that purely data-driven approaches can be confounded by local spectral artefacts, yielding higher false-alarm rates without explicit glitch modelling~\cite{tenorioLearningDetectContinuous2025, cuocoEnhancingGravitationalwaveScience2020}. Furthermore, their reliability is constrained by training priors, making them susceptible to ``out-of-distribution'' artefacts characteristic of the evolving detector state~\cite{mageeMitigatingImpactNoise2024}. This limitation necessitates a robust, unsupervised approach that uses the physical coherence between detectors to distinguish noise from signals without relying on brittle training priors.

Traditional noise mitigation relies on linear subtraction using auxiliary sensors, as established in the feed-forward loops of the O2 and O3 observing runs~\cite{driggersImprovingAstrophysicalParameter2019, davisImprovingSensitivityAdvanced2019, davisSubtractingGlitchesGravitationalwave2022}. While effective for stationary couplings, these methods struggle to capture nonlinear, transient dynamics. Models of nonlinear and/or non-stationary noise couplings, inspired by deep neural networks, have been successful in mitigating low-frequency noise, in particular the 60~Hz line in O3 data~\cite{VajeEtAl2020-MchNnNsOGrvDt, Vaje2022-DMMcLrIGrvDtSns}. Adaptive cancellation of the 60~Hz line using an HMM model has also been demonstrated in O3 data~\cite{KimpEtAl2024-AdCncMPInCnGrvWSHMM}. The late O4 era has seen the deployment of DeepClean, which has evolved from its initial version, a semi-manual de-noising tool using auxiliary sensors~\cite{ormistonNoiseReductionGravitationalwave2020}, to the autonomous Coherence DeepClean pipeline~\cite{reisselCoherenceDeepCleanAutonomous2025}. Concurrently, advanced architectures have emerged, including DeepExtractor specialised in time-domain reconstruction~\cite{dooneyTimedomainReconstructionSignals2025}, and the sample-efficient WaveletNet proposed in~\cite{pimpalkarSampleefficientNonGaussianNoise2026}.

The cataloguing and mitigation of line artefacts (and their harmonics, known as combs) is a well-established component of GW detector characterisation and data quality activities within the LIGO-Virgo-KAGRA Collaboration~\cite{covasIdentificationMitigationNarrow2018, O3-lines, O4a-lines, goetz2026narrowspectralartifactinvestigation}. It is invaluable, not only to data analysts, but also to detector commissioners; identification of the physical cause of a line artefact may enable the removal or suppression of its instrumental source. Line artefacts are typically identified in power spectra averaged over part or all of an observing run, built from Fourier transforms of short ($\sim$ 7200 seconds) segments of GW data. Further manual investigations are generally needed to identify, where possible, the root physical cause of the instrumental disturbance. CW searches have primarily used the known lines catalogues~\cite{O3-lines, O4a-lines} described above as post-hoc vetoes; any candidate signal arising from the initial search who coincides or overlaps in frequency with a known line is discarded. Typically, this known line veto removes many, but not all, CW candidates whose root cause is an instrumental line; further vetoes are then required~\cite{keitelSearchContinuousGravitational2014, zhuNewVetoContinuous2017, JoneEtAl2022-VlCnGrvCnSmcSUDMEPSF} which add person-time and complexity to the post-processing stage of the CW search.

Because an initial CW search evaluates an enormous number of template waveforms, only a small fraction of the most significant templates can be retained for follow-up. Specifically, candidates are selected from the highest-ranked fraction $p$ of detection-statistic values across the search, where $p \ll 1$. Line artefacts can produce large detection-statistic values and thereby occupy this highest-ranked fraction, potentially obscuring weaker but astrophysical candidates. This motivates the cleaning of instrumental artefacts from the input data, so as to avoid contamination of the CW search results. A few CW searches have taken this approach; in~\cite{LIGOVirg2013-EnsAlSrPrGrvWLSD} frequencies coincident with known lines were replaced with simulated Gaussian noise. This type of cleaning, which essentially masks the cleaned frequency bands, necessarily destroys any astrophysical, or part therefore, that may have intersected with the cleaned frequency band. More sophisticated cleaning methods have been successfully used to remove prominent lines such as at 60~Hz~\cite{VajeEtAl2020-MchNnNsOGrvDt, KimpEtAl2024-AdCncMPInCnGrvWSHMM} but have yet to be demonstrated to be effective in mitigating the full spectral forest of line artefacts.

In this paper, we present a generalised spectral line mitigation framework based on adaptive network coherence analysis. This approach overcomes the limitations of frequency masking and avoids the dependence on training priors common in machine learning methods. Our pipeline exploits the physical principle that astrophysical signals maintain coherence across the detector network, whereas instrumental and local environmental artefacts are typically incoherent. We systematically classify spectral features into four morphological categories, ranging from very narrow electronic artefacts to wide mechanical resonances, and apply adaptive coherence criteria derived from the physical characteristics of each class. A high-precision frequency bypass mechanism is employed to preserve global environmental lines, which are characterised by extreme frequency stability. While the framework is designed for an arbitrary number of detectors including Virgo and KAGRA, we demonstrate its efficacy and validate its performance using public data from LIGO Hanford (H1) and Livingston (L1) detectors during the O3 observing run~\cite{O3-data}. We show that this approach effectively removes the spectral forest by rescaling incoherent artefacts to the background noise level, thereby recovering sensitivity in parameter spaces previously discarded by traditional veto methods.

The remainder of this paper is organised as follows. In Section~\ref{sec:methodology}, we detail the methodology, describing the robust baseline fitting, morphological line classification, and the adaptive network coherence logic. In Section~\ref{sec:results}, we present the performance validation using data from the Advanced LIGO O3 observing run, providing a statistical analysis of the cleaning efficacy, its impact on the $\cF$-statistic distribution, and the recovery of simulated CW signals. Finally, Section~\ref{sec:conclusion} presents concluding remarks and future prospects for CW searches.

\section{Methodology}\label{sec:methodology}

We present a generalised pipeline to identify and mitigate non-Gaussian spectral artefacts in GW detector data. The implementation supports an arbitrary number of detector inputs. Baseline fitting and line detection are performed independently for each detector, while the network-coherence stage evaluates all unique detector pairs, requiring \(N_{\mathrm{det}}(N_{\mathrm{det}}-1)/2\) pairwise comparisons for \(N_{\mathrm{det}}\) detectors. The time subsets are independent and can likewise be processed in parallel. The pipeline consists of four primary stages: robust baseline estimation, morphological line detection, adaptive network coherence analysis, and spectral artefact mitigation. The pipeline is available as part of the GW Line Cleaner package~\cite{gw-line-cleaner}.

\subsection{Baseline Fitting and Data Normalisation}

The primary objective of the first stage is to accurately model the detector noise floor. Unlike spectral line artefacts, which exhibit a wide variety of morphologies, the baseline noise floor comprises known fundamental noise sources exhibiting a complex power-law structure. By first robustly fitting the noise baseline, spectral artefacts are then readily identified, i.e. as features that rise about the baseline. Separating the baseline noise floor from the line artefacts in this way greatly aids in robust identification, classification, and mitigation of the latter.

Prior to fitting, Power Spectral Densities (PSDs) are generated from Short Fourier Transforms (SFTs), which are Fourier transforms of short (typically 1800-second) segments of the GW data. No running-median smoothing is applied at this stage, in order to preserve the spectral resolution. At each frequency bin, the PSD estimates from the SFTs within a chosen subset of the O3 data are combined using the harmonic mean,
\begin{equation}
\mathcal{S}_{\mathrm{harm}}(f)
=
\left[
\frac{1}{N_{\mathrm{SFT}}}
\sum_{j=1}^{N_{\mathrm{SFT}}}
\frac{1}{\mathcal{S}_j(f)}
\right]^{-1},
\label{eq:harmonic_mean}
\end{equation}
where \(\mathcal{S}_j(f)\) is the PSD estimated from the \(j\)-th SFT. Compared with the arithmetic mean, the harmonic mean gives less weight to unusually large PSD values occurring in individual SFTs, thereby reducing the influence of isolated spectral excursions on the subset-averaged spectrum. Spectral excesses that persist across the subset remain represented in the combined PSD. The harmonically averaged PSD is then converted to an amplitude spectral density (ASD), on which the subsequent baseline fitting is performed.

A robust iterative algorithm is then employed to fit the noise baseline, as follows:

\begin{enumerate}

    \item Initial Baseline Estimation: A running-median filter with a window size of 151 frequency bins is first applied to the ASD, providing an initial estimate of the underlying noise baseline. For the 1800-second SFTs used here, the frequency-bin spacing is \(1/1800\)~Hz, so the 151-bin window spans approximately 0.084~Hz. The value of 151 bins is a fixed implementation choice used to construct a robust starting estimate, rather than a physically derived line-width threshold; the final baseline is determined by the subsequent peak rejection and iterative Chebyshev fitting. As a nonlinear estimator, the running median is robust against sharp spectral lines that would otherwise bias the initial estimate.

    \item Peak Rejection: The ASD is normalised by the initial baseline, and the most prominent peaks that could bias the subsequent fit are identified. We require a minimum prominence of 3.0, where prominence is defined as the height of a peak relative to its local contour level in the normalised spectrum. For each detected peak, the peak frequency bin and five frequency bins on either side are masked. This wider exclusion accounts for the fact that a spectral line may extend over multiple neighbouring frequency bins and reduces contamination of the subsequent polynomial fit.

    \item Chebyshev Polynomial Fit: A tenth-order Chebyshev polynomial is fitted to the remaining frequency bins in log-frequency--log-ASD space. This representation is motivated by the approximately power-law behaviour of the underlying noise floor, while Chebyshev polynomials provide good numerical stability for higher-order fits. The polynomial degree of 10 was chosen empirically to provide sufficient flexibility to follow the broad structure of the noise floor without fitting local spectral fluctuations.

    \item Iterative Refinement: The fitted polynomial serves as the baseline for the next iteration. At each iteration, logarithmic residuals are calculated as
    $\log_{10}[\mathrm{ASD}(f)/\mathrm{ASD}_{\mathrm{baseline}}(f)]$. The median and Median Absolute Deviation (MAD) of these residuals are computed, and frequency bins with residuals more than 3.0 MADs above the median are excluded before the Chebyshev polynomial is re-fitted. The procedure is repeated for at most 10 iterations, terminating earlier if the MAD falls below $10^{-10}$ or if fewer than 30\% of the original frequency bins remain after outlier rejection.

\end{enumerate}

\subsection{Individual Line Detection}
\label{sec:line-detection}
Following baseline fitting, the ASD is normalised by dividing it by the fitted ASD baseline. The second stage then generates a comprehensive catalogue of statistically significant spectral lines for each individual GW detector. The detection logic consists of three steps:

\begin{enumerate}

    \item Peak Identification: Peaks are identified based on two robust criteria. First, a minimum amplitude ratio is enforced, requiring the peak to exceed the fitted baseline by a factor of 2.5. This threshold is chosen to balance the identification of weaker lines against false alarms arising from statistical fluctuations in the noise floor. Second, a minimum prominence of 0.1 in the normalised ASD is required. Here, prominence measures the height of a peak relative to the reference contour level determined by the surrounding spectrum, thereby rejecting small fluctuations superimposed on larger spectral features.

    \item Boundary Determination: For each identified peak, the algorithm scans outward in frequency from the peak centre. The lower and upper boundaries are defined by the first frequency bins on either side at which the normalised ASD falls to 1.05 or below. This threshold captures the extent of the line while limiting its extension into statistical fluctuations of the surrounding noise floor, which could otherwise lead to the erroneous merging of distinct features.

    \item Logarithmic Clustering: To avoid fragmenting a single broad spectral feature into multiple detected lines, nearby peaks are merged using frequency-dependent clustering criteria. Peaks whose frequency boundaries overlap are always merged. For non-overlapping peaks, merging requires both their peak frequencies and the gap between their boundaries to satisfy logarithmically frequency-dependent tolerances. The allowed relative peak-frequency difference increases from approximately 2\% at 10~Hz to 7\% at 2000~Hz, while the maximum boundary gap increases from approximately 0.5\% to 4\% of the line frequency over the same range. These criteria allow closely related peaks to be grouped while keeping well-separated spectral features distinct.

\end{enumerate}

\subsection{Network Coherence Analysis}
\label{sec:coherence-analysis}

\begin{figure}
    \centering
    \includegraphics[width=1\linewidth]{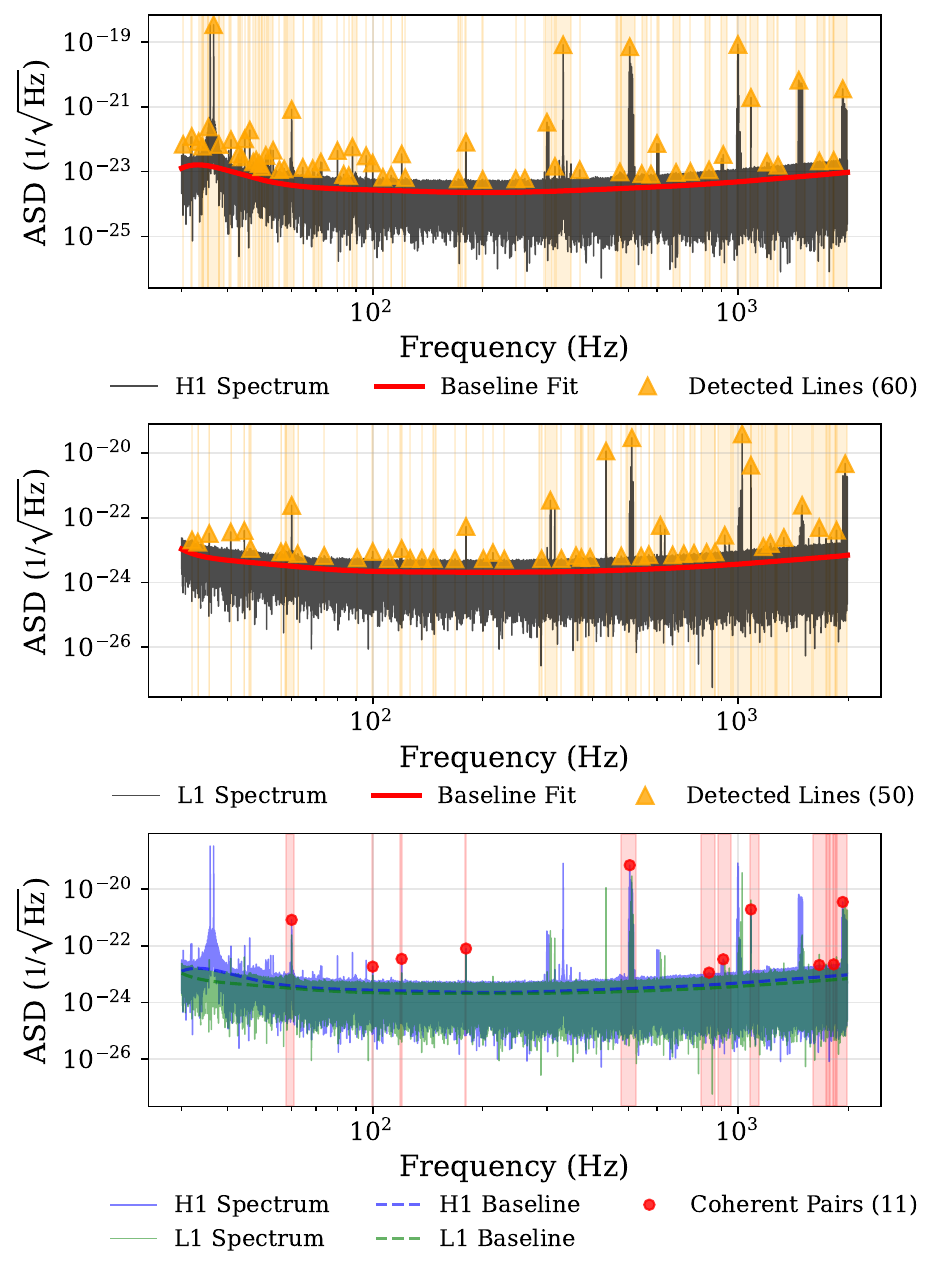}
    \caption{Visualisation of the first three stages of the pipeline applied to Advanced LIGO H1 and L1 data. Top and middle panels: spectral line detection for H1 and L1, respectively, showing the robust baseline fit (red line) and detected lines (orange triangles). Bottom panel: adaptive matching of coherent lines between detectors (red circles).}
    \label{fig:visualisation}
\end{figure}

The third stage of the pipeline compares the identified line catalogues from the second stage across the detector network to distinguish between local instrumental artefacts and global coherent features. The core principle is that astrophysical signals and global environmental noise should appear \emph{coherently} in multiple detectors, whereas most instrumental artefacts are local to one detector and therefore \emph{incoherent}. The analysis proceeds in three steps:

\begin{enumerate}

    \item Morphological Classification: Each detected line is classified into one of four classes based on its bandwidth $W$. We define \emph{very narrow} lines ($W < 5$~mHz) to capture ultra-stable artefacts. The remaining lines are classified as \emph{narrow} ($5\text{~mHz} \le W < 2\text{~Hz}$), \emph{medium} ($2\text{~Hz} \le W \le 10\text{~Hz}$), or \emph{wide} ($W > 10\text{~Hz}$). These thresholds are empirically determined to categorise a wide variety of spectral features ranging from precise electronic harmonics to broad mechanical resonances. The distributions of the line widths identified by the pipeline are discussed in Section~\ref{sec:morphological-classification}.

        \item Adaptive Matching: The criteria for determining whether a feature is coherent across detectors are adapted to the line class identified in the previous step. The quantities used to compare a candidate pair are defined as
    \begin{equation}
    \begin{aligned}
        A &=
        \frac{\mathrm{ASD}_{\mathrm{peak}}}
             {\mathrm{ASD}_{\mathrm{baseline}}}, \\
        d_A &=
        \frac{|A_1-A_2|}
             {\max(A_1,A_2)}, \\
        \Delta f &= |f_1-f_2|, \\
        d_f &=
        \frac{\Delta f}
             {\min(f_1,f_2)} .
    \end{aligned}
    \label{eq:coherence_metrics}
    \end{equation}
    Here, \(A\) is the dimensionless peak-to-baseline ASD ratio evaluated at the line peak, \(d_A\) is the relative amplitude difference, \(\Delta f\) is the absolute frequency separation, and \(d_f\) is the corresponding relative frequency separation.

    For very narrow lines, we apply a strict absolute frequency tolerance of \(1/3600\)~Hz (\(\approx 0.28\)~mHz). This criterion is intended to identify ultra-stable frequency coincidences across detectors. In the implementation, any candidate with \(|\Delta f|<0.1\)~Hz that has already satisfied its class-specific frequency criterion is treated as a high-precision coincidence and is not rejected by the amplitude or boundary-overlap cuts. Every very-narrow pair satisfying the \(1/3600\)~Hz requirement therefore also satisfies this high-precision condition. Consequently, amplitude does not act as an eligibility veto for very-narrow candidates: eligibility is determined by the stringent frequency criterion, while \(d_A\) can still affect the ranking if more than one frequency-compatible candidate is available.

    For narrow lines, a relative frequency tolerance of 0.1\% is applied. For medium lines, the tolerance is 0.5\%, together with a minimum 5\% overlap of the line-frequency intervals. For wide lines, the frequency tolerance is 2\% and the minimum overlap is 10\%. Outside the high-precision condition described above, narrow-line candidates with \(d_A>0.80\) are rejected. The corresponding amplitude threshold is 1.0 for medium and wide candidates; for positive amplitudes this does not normally act as an additional rejection cut, although amplitude consistency still contributes to the candidate-ranking score below. For pairs with different morphological classifications, the matching criteria are selected explicitly according to the pair: if either line is very narrow, the very-narrow criteria are used; pairs in which both lines are narrow or both are wide use the corresponding narrow or wide criteria, respectively; all other mixed-morphology pairs are evaluated using the medium-line criteria.

    \item Scoring and Grouping: Candidate pairs that pass the class-specific eligibility criteria are ranked to resolve cases in which more than one match is possible. Let \(O\) denote the intersection width of the two line-frequency intervals divided by their union width. The candidate score is
    \begin{equation}
    S =
    \begin{cases}
        3\,\Delta f/\delta_{\mathrm{VN}} + 0.5\,d_A,
        & \text{very narrow},\\
        3\,d_f + 0.5\,d_A,
        & \text{narrow},\\
        2\,d_f + 0.5\,d_A + (1-O),
        & \text{medium},\\
        d_f + 0.5\,d_A + 2(1-O),
        & \text{wide},
    \end{cases}
    \label{eq:coherence_score}
    \end{equation}
    where \(\delta_{\mathrm{VN}}=1/3600\)~Hz. Lower values of \(S\) indicate better matches. Within each detector pair, the algorithm proceeds through the detected lines in one detector and retains the lowest-scoring eligible unused candidate in the other detector. The resulting pairwise matches are subsequently consolidated across the network; pairwise matches whose representative frequencies differ by less than 1\% are assigned to the same coherent group.

    A feature is flagged as \emph{coherent} if it appears in at least two detectors within the network. Lines identified as coherent are preserved to protect potential astrophysical signals. Lines appearing in only a single detector are classified as \emph{incoherent} noise and targeted for mitigation in the subsequent cleaning stage.

\end{enumerate}

We do not pre-clean catalogue entries or remove very-narrow features before applying the coherence analysis. Such a step would impose an external classification and would remove that part of the detected-line population from an end-to-end test of the network criteria. Instead, all detected features are processed by the same coherence stage: detector-specific very-narrow features remain eligible for mitigation, while sufficiently frequency-coincident features are retained. This also avoids treating bandwidth alone as evidence of instrumental origin.

An illustration of the first three stages of the pipeline is presented in Figure~\ref{fig:visualisation}. The plots demonstrate the pipeline's robustness in identifying coherent groups within the intrinsic statistical fluctuations of the raw detector data, successfully distinguishing them from a dense noise floor.

\subsection{Spectral Artefact Mitigation}

The mitigation of spectral artefacts is performed in the fourth stage of the pipeline. In contrast to cleaning methods which discard contaminated data entirely, by either zeroing out specific frequency bands or replacing them with simulated Gaussian noise~\cite{LIGOVirg2013-EnsAlSrPrGrvWLSD}, we employ a frequency-domain substitution strategy that surgically suppresses artefacts without creating gaps in the data or losing information from the surrounding frequency bins.

Specifically, spectral lines classified as incoherent are directly rescaled to the level of the fitted noise baseline. For every frequency bin $k$ between the lower and upper line boundaries defined in Section~\ref{sec:line-detection}, where the normalised ASD returns to 1.05 or below, we desire that the PSD $\mathcal{S}(f_k)$ should equal the fitted noise baseline, $\mathcal{S}_{\text{baseline}}(f_k)$. To achieve this, the complex-valued SFT coefficients $\tilde{x}^{\text{original}}_k$ of the original GW data are adjusted according to:
\begin{equation}
    \label{eq:cleaning}
    \tilde{x}^{\text{cleaned}}_k = \tilde{x}^{\text{original}}_k \sqrt{\frac{\mathcal{S}_{\text{baseline}}(f_k)}{\mathcal{S}_{\text{original}}(f_k)}},
\end{equation}
where $\mathcal{S}_{\text{original}}(f_k)$ represents the original PSD before the pipeline is applied, and $\tilde{x}^{\text{cleaned}}_k$ are the complex values of the now-cleaned SFTs. This real-valued rescaling suppresses the excess power associated with an incoherent line towards the fitted noise baseline. Because the rescaling factor is real and positive, it does not introduce an additional phase rotation to the observed complex SFT coefficient. However, if a strong instrumental line dominates a frequency bin, the phase of that coefficient may itself be dominated by the artefact; the procedure therefore does not recover the phase of a weaker CW contribution hidden beneath it. A CW contribution within a cleaned frequency bin is attenuated by the same rescaling factor as the full complex coefficient. Frequency bins corresponding to coherent lines remain unmodified to conservatively protect potential CW signals.

Thus, Figure~\ref{fig:visualisation} provides a visual summary of the first three stages of the pipeline, while Equation~(\ref{eq:cleaning}) gives the explicit transformation applied in the fourth, mitigation stage.

\section{Results}\label{sec:results}

We applied the adaptive coherence mitigation pipeline to the Advanced LIGO O3 data~\cite{O3-data}. We use as input the standard 1800-second SFTs generated at times listed in~\cite{O3-SFT-segments}. To assess the robustness of the method against different PSD-averaging spans and artefact persistence timescales, the O3 SFTs are partitioned into subsets spanning 1 day, 3 days, and 5 days, specifically selected to align with the typical coherent segment lengths employed in standard semi-coherent search strategies. Here, 1, 3, and 5 days denote the nominal wall-clock spans of the subset windows (86400, 259200, and 432000~s), rather than accumulated good-data live time. Only the available 1800-second SFTs within each window are analysed, so data gaps can reduce the actual live time contributing to a subset. Rather than relying on a single continuous observation, this analysis covers a statistically significant sample of subsets distributed across the observing run to marginalise over the non-stationary evolution of the detector noise floor.

To systematically quantify the pipeline's efficacy in noise mitigation and its safety regarding astrophysical CW signals, we evaluate its performance using the following metrics. The metrics defined below are computed as ensemble statistics across the 1-day, 3-day, and 5-day subsets. We use 95\% bootstrap confidence intervals to quantify subset-to-subset sampling variability without assuming a Gaussian sampling distribution for these statistics. Further details of the following evaluations are given in the following sections.

\begin{enumerate}

    \item Spectral Background Consistency (Section~\ref{sec:spectr-backgr-cons}): As an initial qualitative check, we compare the Amplitude Spectral Density (ASD) and $\cF$-statistic spectra before and after mitigation. This visual comparison serves to verify that incoherent artefacts are suppressed to the baseline noise level and that the fundamental noise floor remains intact without introducing spectral discontinuities.

      The ASDs are simply the square root of the previously computed PSDs. The $\cF$-statistic values are computed, using the LALSuite package~\cite{lalsuite,swiglal}, as a function of frequency, coherently over each 1-day, 3-day, and 5-day subset. For this test, we compute the $\cF$-statistic for an isolated CW signal (i.e. no modulation from a binary companion), at an arbitrary sky position (right ascension and declination both zero, however Doppler modulation is still included), and with zero frequency derivative parameters. The $\cF$-statistic values thus computed represent a first estimate of the effect of the line cleaning on a CW search.

    \item Global Cleaning Statistics (Section~\ref{sec:glob-clean-stat}): To quantify the cleaning efficiency and invasiveness, we compute aggregate line statistics. First, we track the reduction in the total count of spectral lines, separated into incoherent (cleaned) and coherent (retained) populations to confirm the targeting efficiency. Second, we compute the \emph{cleaned bandwidth fraction}, defined as the percentage of the analysis band (20-2000~Hz) modified by the mitigation mask. A low fraction indicates high spectral selectivity.

    \item Morphological Classification (Section~\ref{sec:morphological-classification}): To compare the morphology of the retained and removed features, we examine the run-normalised frequency and line-width distributions of the coherent and incoherent populations. These distributions show where the two populations differ and where they overlap, providing a direct assessment of bandwidth as one component of the adaptive coherence criteria.

    \item Statistical Verification (Section~\ref{sec:stat-verif}): This validation is performed using a Quantile-Quantile (Q-Q) analysis of the $\cF$-statistic, partitioned into two regimes. For frequency bins outside coherent groups, we calculate the tail slope $s$ via a linear fit to the upper quantiles ($q \in [0.90, 0.99]$) of the Q-Q plot, where $s \to 0$ indicates effective artefact suppression. Conversely, for bins inside coherent groups, we define the tail ratio $r$ as the median ratio of $2\cF_{\text{after}} / 2\cF_{\text{before}}$ for the same upper quantiles. A value of $r \approx 1$ confirms that the coherent power is invariant under cleaning, ensuring that potential astrophysical signals are preserved.

    \item Recovery of Simulated Signal Injections (Sections~\ref{sec:hw-inj} and~\ref{sec:sw-inj}): Here we verify that the cleaning does not impede the recovery of simulated CW signals outside of the cleaned bands. To do this, we use two classes of simulated signals: the O3 hardware injections, simulated CW signals with known parameters embedded directly in the O3 data; and software injections, CW signals simulated in software with parameters drawn from a wide parameter space of frequencies, frequency derivatives, sky positions, and amplitudes.

    \item Comparison to O3 Lines List (Section~\ref{sec:comparison-o3-lines}): Finally, we compare and contract the lines typically found by the pipeline against the O3 lines list produced by the LIGO-Virgo-KAGRA Collaboration~\cite{O3-lines}.

\end{enumerate}

To visualise the spectral impact, Figure~\ref{fig:spectral_impact} presents an illustrative 5-day subset, corresponding to the longest PSD-averaging span considered in this work. In contrast, the global cleaning statistics (Fig.~\ref{fig:global_cleaning_stats}), morphological classifications (Fig.~\ref{fig:morphological_dists}), and rigorous statistical verification (Fig.~\ref{fig:stat_verification}) are aggregated across the full ensemble of 1-day, 3-day, and 5-day subsets to ensure the statistical generality of the results. Specifically, we report the ensemble mean for spectral line counts and bandwidths shown in Fig.~\ref{fig:global_cleaning_stats} to quantify total cleaning impact, while employing the ensemble median for the Quantile-Quantile (Q-Q) analysis shown in Fig.~\ref{fig:stat_verification} to robustly characterise the typical noise suppression performance against outliers.

\subsection{Spectral Background Consistency}\label{sec:spectr-backgr-cons}

\begin{figure}
    \centering
    \begin{subfigure}{\linewidth}
        \centering        \includegraphics[width=1\linewidth]{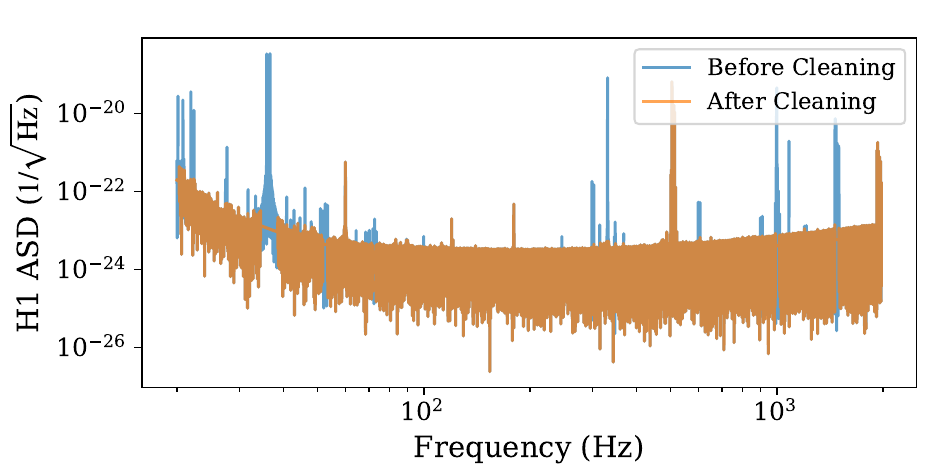}
    \end{subfigure}
    \begin{subfigure}{\linewidth}
        \centering        \includegraphics[width=1\linewidth]{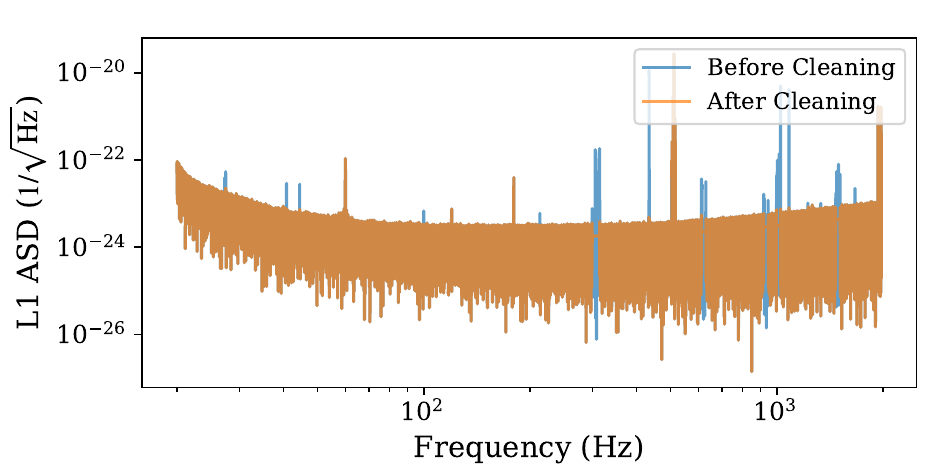}
    \end{subfigure}
    \begin{subfigure}{\linewidth}
        \centering        \includegraphics[width=1\linewidth]{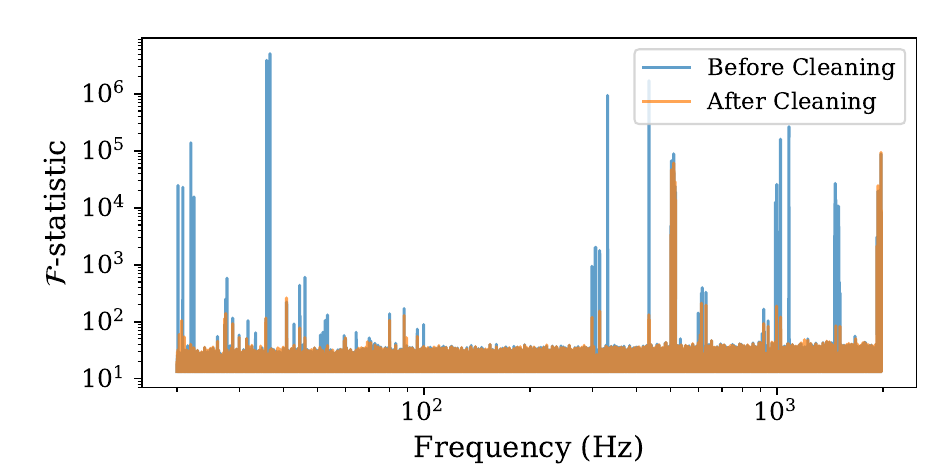}
    \end{subfigure}
    \caption{Spectral impact of adaptive coherence mitigation on a 5-day subset of Advanced LIGO O3 data. Top and middle panels: ASDs for H1 and L1, respectively. Bottom panel: $\cF$-statistic values. All plots show results before (blue) and after (orange) cleaning.}
    \label{fig:spectral_impact}
\end{figure}

The spectral impact of the mitigation is visualised in Figure~\ref{fig:spectral_impact}, which compares the ASD and the detection statistic $2\cF$ before and after cleaning. The raw data before cleaning is dominated by a dense spectral forest, with $\cF$-statistic values frequently exceeding $10^3$.\footnote{The probability of an $\cF$-statistic value greater than $10^3$ in purely Gaussian noise is $\sim 10^{-215}$.} After the application of the cleaning, the dense noise lines are effectively suppressed. The cleaned spectrum tracks the underlying continuum noise floor, verifying that the algorithm successfully identifies and rescales non-Gaussian outliers. Crucially, the mitigation does not introduce spectral discontinuities. The cleaned ASD tracks the underlying continuum noise level, represented by the fitted baseline, verifying that incoherent artefacts are rescaled to the fundamental sensitivity limit governed by quantum and thermal noise. Furthermore, prominent spectral features  (e.g., at $\sim 510$~Hz) remain visible in the cleaned data. These correspond to globally coherent signals satisfying the coherence criteria, visually validating the signal safety mechanism.

\subsection{Global Cleaning Statistics}\label{sec:glob-clean-stat}

\begin{figure}
    \centering
    \begin{subfigure}{\linewidth}
        \centering        \includegraphics[width=1\linewidth]{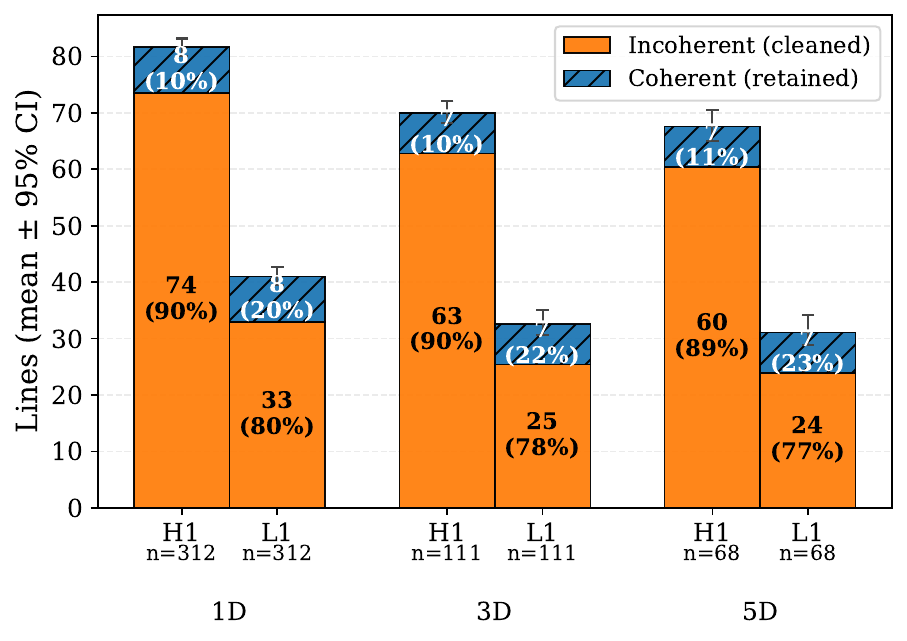}
    \end{subfigure}
    \begin{subfigure}{\linewidth}
        \centering        \includegraphics[width=1\linewidth]{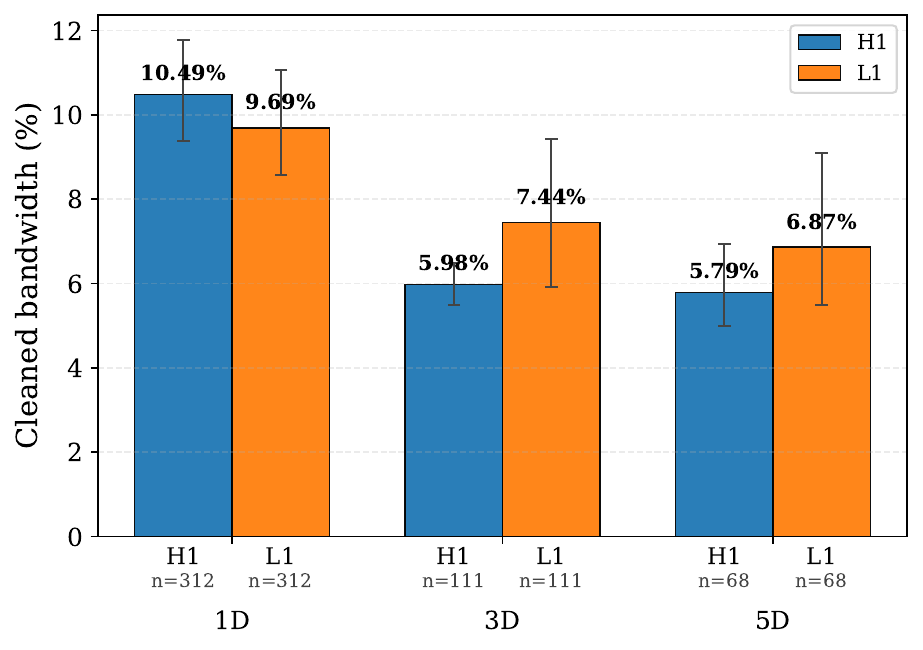}
    \end{subfigure}
    \caption{Global cleaning statistics versus subset time-span, aggregated across all subsets ($n=312$, $111$, and $68$ for 1, 3, and 5 days, respectively). Top panel: ensemble-mean counts of incoherent (cleaned) and coherent (retained) spectral lines; the full height of each stacked bar represents the total ensemble mean. Numerical annotations are rounded to the nearest integer for display, whereas the bar heights are computed from the unrounded ensemble means. Bottom panel: percentage of the analysis band (20--2000~Hz) modified by the mitigation. Error bars represent 95\% bootstrap confidence intervals.}
    \label{fig:global_cleaning_stats}
\end{figure}

We quantify the cleaning efficiency using the spectral line counts and bandwidth fractions presented in Figure~\ref{fig:global_cleaning_stats}. Summarising the performance across the different subset time-spans, these statistics represent the ensemble mean computed from $n=312$ (1-day), $n=111$ (3-day), and $n=68$ (5-day) independent subsets of the O3 data. For each detector and subset, the total detected-line population is partitioned into disjoint incoherent (cleaned) and coherent (retained) populations. Line counts quoted in the text and annotated in the figure are ensemble means rounded to the nearest integer for readability; the stacked-bar heights and confidence intervals are calculated from the corresponding unrounded values. The same line-detection criteria are applied to H1 and L1, so the larger number of detected features in H1 is not a consequence of detector-dependent thresholds in the algorithm. We do not attempt to attribute this detector-to-detector difference to a specific instrumental cause in the present work. This comprehensive comparison allows us to track the evolution of spectral artefacts and the algorithm's selectivity as the subset time-span increases.

Focusing on the 5-day subsets, as shown in the upper panel, the pipeline identified and removed 60 incoherent lines (89\% of the total) in H1 and 24 lines (77\%) in L1, while preserving 7 coherent features in each detector. This high rejection rate is achieved with minimal data loss. As shown in the lower panel, the cleaned bandwidth fraction is 5.79\% for H1 and 6.87\% for L1. This indicates that the algorithm achieves high spectral selectivity, modifying less than 7\% of the analysis band even in the presence of dense artefact accumulation.

The evolution of these metrics with subset time-span reflects a change in the line population retained above the detection threshold. The total number of detected lines decreases slightly from 82 to 68 in H1 and from 41 to 31 in L1 as the subset time-span lengthens, consistent with some short-duration or non-stationary excesses being diluted when the PSD is averaged over a longer window. The cleaned bandwidth fraction likewise decreases from 10.49\% (1-day) to 5.79\% (5-day) for H1 and from 9.69\% (1-day) to 6.87\% (5-day) for L1. Because all three analyses use the same 1800-second SFTs, the native SFT frequency-bin spacing is unchanged; the bandwidth decrease reflects the changing population and extent of features selected for mitigation rather than a change in the underlying SFT frequency resolution.

\subsection{Morphological Classification}\label{sec:morphological-classification}

\begin{figure}
    \centering
    \includegraphics[width=1\linewidth]{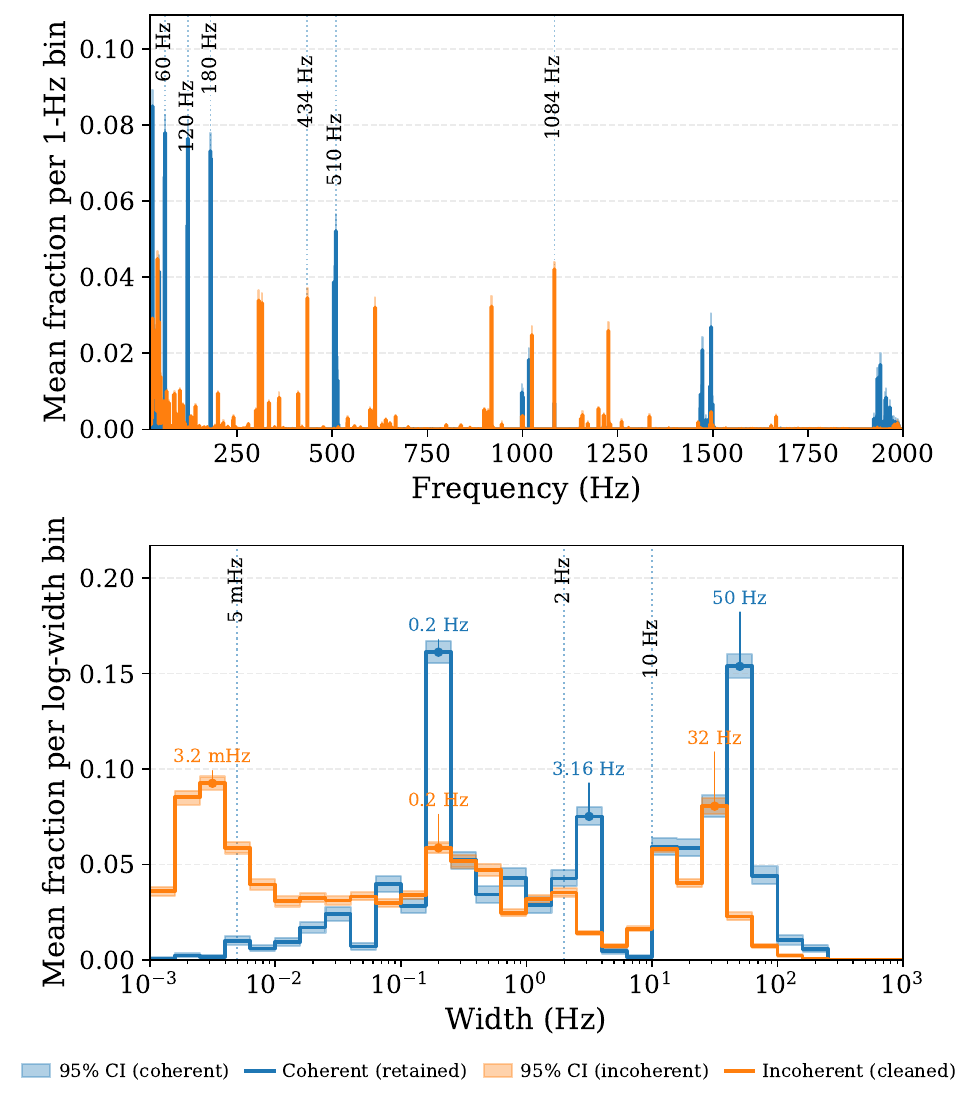}
    \caption{Morphological distributions of coherent (retained; blue) and incoherent (cleaned; orange) spectral features identified across the $491$ 1-day, 3-day, and 5-day O3 subsets. For each detector-subset sample and each population separately, the histogram is normalised by the number of features in that population, so that the bin fractions sum to unity. The plotted curves show the ensemble mean over detector-subset samples containing at least one feature of the corresponding population, with shaded regions indicating 95\% bootstrap intervals. Top panel: mean fraction per 1-Hz frequency bin. Selected frequencies at 60, 120, 180, 434, 510, and 1084~Hz are labelled for reference. Bottom panel: mean fraction per logarithmic line-width bin, using 30 logarithmically spaced bins spanning the full observed positive line-width range. Vertical dotted lines mark the morphology-class boundaries at 5~mHz, 2~Hz, and 10~Hz; labelled points indicate prominent modes of the coherent and incoherent width distributions.}
    \label{fig:morphological_dists}
\end{figure}

To compare the morphology of the retained and removed populations, we examine their frequency and line-width distributions in Figure~\ref{fig:morphological_dists}. Each detector-subset histogram is normalised separately for the coherent and incoherent populations before the ensemble average is calculated. This representation emphasises the shape of the distributions rather than the different absolute numbers of coherent and incoherent features.

The frequency distributions show a clear difference between the two populations. The coherent population contains several sharp, localised features, including prominent peaks at 60, 120, and 180~Hz (the U.S. mains power harmonics), and near 510~Hz. In contrast, the incoherent population is more broadly distributed across the analysis band, with numerous detector-specific features at other frequencies. The labelled frequencies at 434 and 1084~Hz provide representative examples of prominent structure outside the 60-Hz harmonic sequence. We do not assign a specific physical origin to every labelled feature here; the purpose of this comparison is to characterise the frequency morphology of the populations selected by the coherence analysis.

The line-width distributions are clearly multimodal and show substantial overlap between the coherent and incoherent populations. The incoherent population has a prominent very-narrow mode near 3.2~mHz, below the 5~mHz classification boundary, together with additional modes near 0.2~Hz and 32~Hz. The coherent population also contains a mode near 0.2~Hz, as well as broader modes near 3.16~Hz and 50~Hz. These values correspond to the centres of prominent logarithmic histogram bins and should therefore be interpreted as approximate characteristic widths rather than fitted line-width parameters. The overlap between the two populations shows that bandwidth alone does not provide a complete separation between coherent and incoherent features. Instead, it provides one morphological input to the adaptive matching criteria, together with frequency consistency and the other coherence measures described in Section~\ref{sec:coherence-analysis}.

\subsection{Statistical Verification}\label{sec:stat-verif}

\begin{figure}
    \centering    \includegraphics[width=1\linewidth]{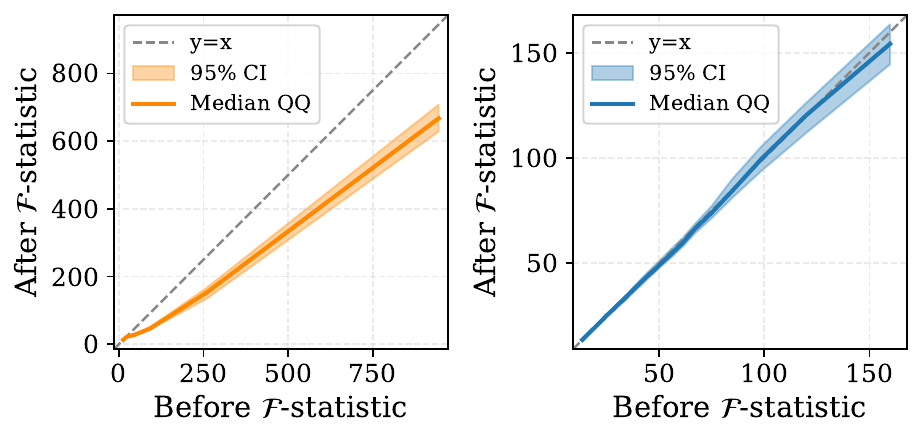}
    \caption{Q-Q analysis of the $\cF$-statistic for all subsets ($n=491$). Left panel: for frequency bins associated with incoherent lines and the background (orange). Right panel: for frequency bins associated with coherent features (blue). Shaded regions indicate 95\% bootstrap confidence intervals.}
    \label{fig:stat_verification}
\end{figure}

The rigorous verification of signal safety and noise suppression is performed using the Q-Q analysis of the $\cF$-statistic presented in Figure~\ref{fig:stat_verification}. The metrics reported below represent the ensemble median derived from the full dataset (i.e. across all 1-day, 3-day, and 5-day subsets) to robustly characterise the typical performance. For frequency bins identified as coherent, the Q-Q plot follows the $y=x$ diagonal almost perfectly, yielding a tail slope of $s = 0.984$ (95\% CI: $[0.94, 1.04]$) and a tail ratio of $r = 0.982$ (95\% CI: $[0.97, 1.00]$). These values are statistically indistinguishable from unity, confirming that the amplitude and statistical distribution of coherent signals are preserved with high fidelity. The adherence to the diagonal confirms that the mitigation strategy acts as a safe veto for non-Gaussian outliers without introducing systematic bias to the signal population.

For the incoherent lines and background, the distribution shows a significant reduction in outlier power. The observed tail slope of $s = 0.688$ (95\% CI: $[0.65, 0.73]$) reflects the systematic damping of non-Gaussian artefacts. While a slope of zero would imply complete statistical independence, the observed value of $s \approx 0.69$ indicates a conservative rescaling strategy. This behaviour confirms that the algorithm effectively suppresses artefact amplitudes by orders of magnitude, restoring the noise floor dominance without imposing artificial nulls or discontinuities that could compromise data integrity.

\subsection{Recovery of O3 Hardware Injections}\label{sec:hw-inj}

\begin{figure}
  \centering
  \includegraphics[width=\linewidth]{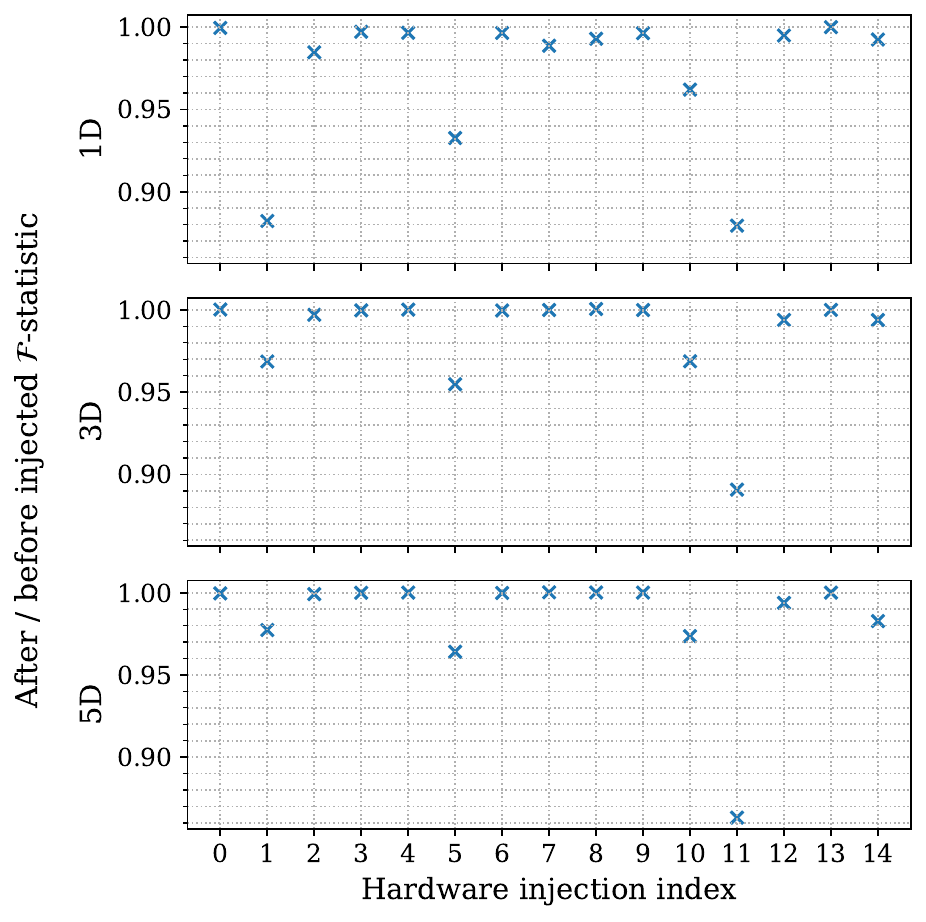}
  \caption{Recovery of the O3 hardware injections. Ratio of the $\cF$-statistic, after/before cleaning, computed at the parameters of hardware injections 0--14. Panels are for (top to bottom) the 1, 3, and 5 day subsets respectively.}
  \label{fig:hw-inj}
\end{figure}

During the O3 run, simulated CW signals were injected into the gravitational-wave strain data by physically moving the test masses of the interferometer~\cite{BiweEtAl2017-VldGrvDtAdLHInS}. Simulated signals generated in this way are known as ``hardware injections'' -- as opposed to ``software injections'' (Section~\ref{sec:sw-inj}) where such signals are added as part of a data analysis pipeline. The O3 run contains 18 CW hardware injections, whose parameters can be found at~\cite{O3-hw-inj}.

We compute the $\cF$-statistic at the parameters of the CW hardware injections before and after the cleaning is applied. This is repeated for the 1-day, 3-day, and 5-day subsets. We consider only the first 15 hardware injections (indices 0--14), all of which are for isolated CW signals (i.e. no modulation from a binary companion). The frequency of injection 15 ($\sim 2991$~Hz) is outside the maximum frequency (2000 Hz) of the standard 1800-second SFTs. Injections 16 and 17 include binary modulation from a companion star; the choice of orbital parameters causes the CW signal frequency to change more rapidly with time, which consequently would require SFTs with time baselines shorter than 1800 seconds in order to compute the $\cF$-statistic.

Figure~\ref{fig:hw-inj} demonstrates the recovery of the hardware injections after cleaning. Typically, for each injection, the $\cF$-statistic computed after cleaning is reduced by less than a few percent, relative to the $\cF$-statistic computed before cleaning. This reduction is consistent with small changes in the PSD of the SFTs due to cleaning, which is likely to have a small effect on the internal normalisation of the data by the PSD performed during the $\cF$-statistic computation. Injections 1, 5, 10, and especially 11 typically suffer a greater reduction in the post-cleaning $\cF$-statistic; injections 1, 5, 10 are typically reduced by $\lesssim 5$\% (except in the 1-day subset where the reduction is $\lesssim 12$\%), while injection 11 is typically reduced by 10--15\%. The frequencies of these injections likely coincide with spectral regions where line artifacts are more prevalent: for example, injection 11 ($\sim 31.4$~Hz) is located at the typically noiser low-frequency end of the spectrum (see Figure~\ref{fig:spectral_impact}). In any case, a reduction of at most $\lesssim 15$\% in $\cF$-statistic is modest: CW searches typically lose $\sim 10$--50\% in signal-to-noise ratio due to template bank effects~\cite{wetteFlatParameterspaceMetric2013}. We conclude that the cleaning has minimal effect on hardware injection recovery.

\subsection{Recovery of Software Injections}\label{sec:sw-inj}

\begin{figure}
  \centering
  \includegraphics[width=\linewidth]{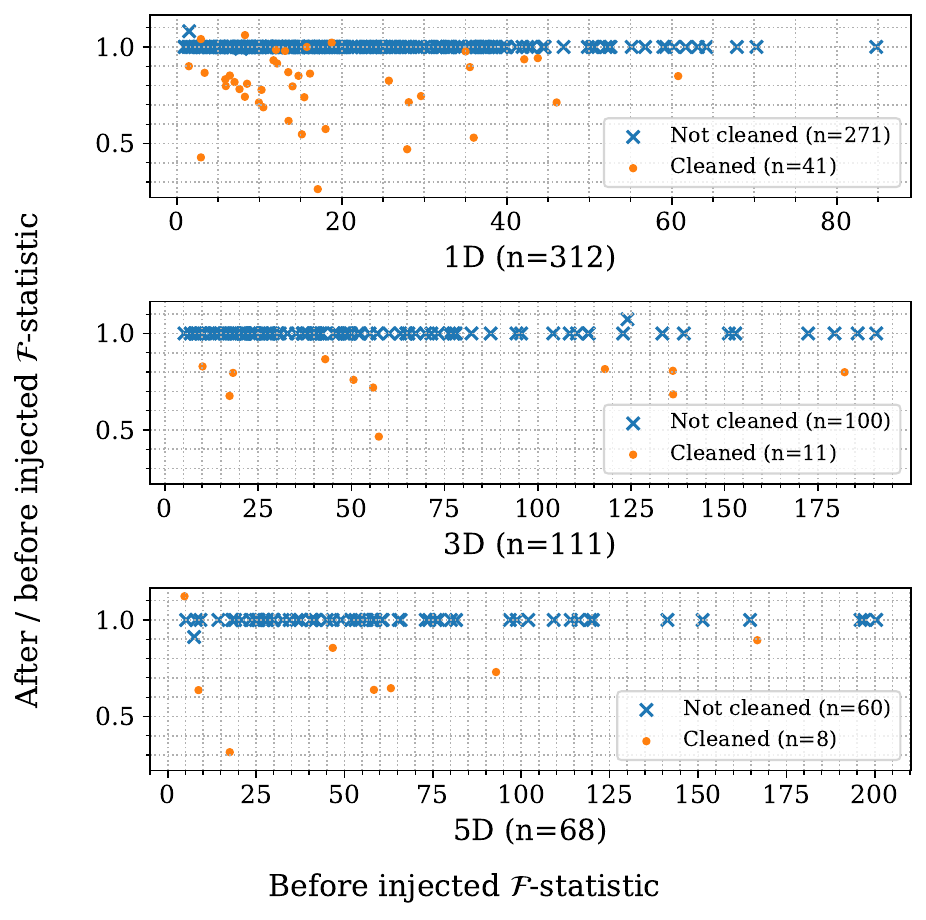}
  \caption{Recovery of software injections. Ratio of the $\cF$-statistic, after/before cleaning, computed at the parameters of software injections added to the O3 SFTs. Panels are for (top to bottom) the 1, 3, and 5 day subsets respectively. Blue crosses: injections that did not overlap in frequency with a cleaned band. Orange dots: injections that did overlap in frequency with a cleaned band. Values of $n$ give the number of (cleaned, not cleaned) injections for each subset.}
  \label{fig:cw-inj}
\end{figure}

To further demonstrate the recovery of CW signals following the cleaning pipeline, we inject a single signal simulated in software into each 1-, 3-, and 5-day subset, totalling 491 injections across all subsets. The $\cF$-statistic is computed at the parameters of the signal, before and after the cleaning is applied. The parameters of the signals are sampled at random as follows.

The frequency $f_{\text{inj}}$ is chosen log-uniformly from the range 50--1950~Hz; note that there is no attempt made to avoid line artefacts in choosing $f_{\text{inj}}$. Each signal has a single frequency derivative parameter (first spin down) chosen log-uniformly from $-10^{-8}$ to $-10^{-12}$ Hz/s. The sky position is a randomly-chosen point on a 2-sphere. No binary orbit parameters are included. The hypothetical neutron star radiating the simulated signal is assumed to be a tri-axial ellipsoid (i.e. its three principal moments of inertia are all different). The signal strength is determined by the strain amplitude $h_0$ and the angle $\cos\iota$ (being the angle of the angular momentum of the neutron star to the line of sight). We select $h_0$ according to
\begin{equation}
  h_0 = \frac{ \sqrt{\mathcal{S}(f_{\text{inj}}) \cdot \text{Hz}} }{ \mathcal{D} } \,,
\end{equation}
where $\mathcal{S}(f_{\text{inj}})$ is the PSD at the injection frequency, and the so-called ``sensitivity depth'' $\mathcal{D}$ \cite{BehnEtAl2015-PstMtUSCnGrvSGlC, DreiEtAl2018-FAcSnsEsCntSr, wetteSearchesContinuousGravitational2023} is chosen log-uniformly from 25--50; in short, signal amplitudes are chosen to be 25 to 50 times weaker than the surrounding noise ASD. Two remaining parameters of the signal -- the polarisation angle $\psi$, and the initial phase $\phi_0$ -- are chosen uniformly from 0--$2\pi$.

Figure~\ref{fig:cw-inj} shows the ratio of the computed $\cF$-statistics (after vs. before cleaning) for $n=312$ signals injected into the 1-day subsets, $n=111$ signals injected into the 3-day subsets, and $n=68$ signals injected into the 5-day subsets. By random chance, 41 injections overlapped in frequency with a cleaned band in the 1-day subsets; 11 injections in the 3-day subsets; and 8 in the 5-day subsets. The remaining injections did not overlap with any cleaned band. It is clear that the cleaning has no impact on injections that do not overlap with any cleaned band; these injections are recovered with no reduction in the $\cF$-statistic. In contrast, the injections overlapping with cleaned bands are understandably impacted. Except for 5 injections, however, that impact is limited to a reduction in $\cF$-statistic of $\lesssim 50$\%, comparable to the maximum loss typically sustained due to template bank effects~\cite{wetteFlatParameterspaceMetric2013}.

In summary, we conclude that the cleaning has no impact on CW signals outside cleaned bands. While CW signals within cleaned bands are impacted, that impact may not be so great as to prohibit detection. We note that, without cleaning, the detection of CW signals close to line artefacts is challenges due to e.g. the saturation of candidate top-lists due to the artefacts obscuring any signal present. It is possible that the cleaning, by removing strong artefacts, may aid in detecting nearby signals by making them more prominent (compared to the cleaned noise), even if their signal-to-noise ratio is impacted.

\subsection{Comparison to O3 Lines List}\label{sec:comparison-o3-lines}

\begin{table}
  \centering
  \begin{tabular}{lrrr}
    \hline
    \hline
    H1               & Found in O3 lines list & Not found in O3 lines list & Total \\
    In this work     & 26 (43\% of total)     & 34 (57\% of total)         & 60    \\
    \hline
    H1               & Found in this work     & Not found in this work     & Total \\
    In O3 lines list & 342 (67\% of total)    & 171 (33\% of total)        & 513   \\
    \hline
    \hline
    L1               & Found in O3 lines list & Not found in O3 lines list & Total \\
    In this work     & 13 (26\% of total)     & 37 (74\% of total)         & 50    \\
    \hline
    L1               & Found in this work     & Not found in this work     & Total \\
    In O3 lines list & 178 (83\% of total)    & 37 (17\% of total)         & 215   \\
    \hline
    \hline
    H1--L1           & Found in O3 lines list & Not found in O3 lines list & Total \\
    In this work     & 2 (18\% of total)      & 9 (82\% of total)          & 11    \\
    \hline
    H1--L1           & Found in this work     & Not found in this work     & Total \\
    In O3 lines list & 2 (8\% of total)       & 22 (92\% of total)         & 24    \\
    \hline
    \hline
  \end{tabular}
  \caption{Comparison of lines identified in the 1-day O3 subset beginning at GPS time 1238198400 with the O3 lines list~\cite{O3-lines}. For H1, L1, and coherent (H1--L1) lines, each section of the table shows: the number of the lines found in this work, and either found or not found in the O3 lines list; and conversely the number of lines in the O3 lines list, and either found or not found in this work.}
  \label{tab:comparison}
\end{table}

Table~\ref{tab:comparison} shows an illustrative comparison between the lines found in the 1-day O3 subset beginning at GPS time 1238198400, also used in Figure~\ref{fig:visualisation}, and the O3 lines list~\cite{O3-lines}. Lines from one source are considered to be found in the other source if their frequency intervals intersect. In this comparison, a line identified by our pipeline but not found in the O3 lines list means only that its frequency interval does not intersect an entry in the O3 lines list. Such an unmatched feature should not, on the basis of this comparison alone, be interpreted as either a false detection or a newly identified instrumental line. For simplicity, we do not compare to comb features listed in the O3 lines list. As the O3 lines list does not supply an explicit list of lines coincident between H1 and L1, we generate such a list (Table~\ref{tab:coherent-lines}) from the H1 lines in the O3 lines list which intersect with lines in L1 (and vice versa); in this combined list, adjacent overlapping frequency bands are coalesced.

Typically, the pipeline identifies some fraction of the lines in the O3 lines list. We find that $\sim 43\%$ of H1 lines, $\sim 26\%$ of L1 lines, and $\sim 18\%$ of coherent (H1--L1) lines found by the pipeline are also found in the O3 lines list. Conversely, while the pipeline finds high fractions of the O3 lines list lines ($\sim 67\%$ of H1 lines, $\sim 83\%$ of L1 lines), it identifies only $\sim 8\%$ of the lines coincident between the H1 and L1 lists of the O3 lines list.

The difference in performance between the pipeline and the O3 lines list is consistent with methodological differences between the two analyses, most notably the PSD-averaging time-scale. Here, we average the PSDs over at most 5 days, whereas the O3 lines list is constructed from spectra accumulated over substantially longer portions of the observing run. Longer averaging can increase the visibility of weak persistent features that may not rise sufficiently above the local baseline in the shorter subsets considered here. Differences in the spectral construction and grouping of nearby features can also cause the same underlying spectral structure to be represented differently in the two analyses. We therefore expect the 1--5-day analysis to recover fewer weak persistent lines and, in some cases, to group nearby features differently from the O3 lines list.

Table~\ref{tab:coherent-lines} lists the coherent lines typically found by this pipeline, and in the O3 lines list. The pipeline generally identifies particularly loud artefacts (e.g. the U.S. mains power harmonics at 60, 120, and 180 Hz), and broader bands contains either a single loud artefact or a cluster or many artefacts (e.g. 819.5--832.3 Hz, 1825.1--1858.1 Hz). The lines identified in the O3 lines list are generally narrow, and more evenly spread in frequency, suggesting that the longer spectral averaging used to construct the list is identifying more weaker but persistent artefacts across the band.

Overall, the results presented in this section demonstrate that the pipeline can robustly identify and mitigate, in a fully automated algorithm, a useful fraction of the line artefacts typically found on GW data, resulting in reduced spectral contamination and fewer false positive CW candidates identified by the $\cF$-statistic.

\section{Conclusion and Outlook}\label{sec:conclusion}

In this work, we demonstrate a robust framework for mitigating the spectral forest in GW data. By exploiting the distinct morphological behaviours of global potentially astrophysical CW signals versus local artefacts, our adaptive coherence pipeline achieves high selectivity without relying on auxiliary channel regression or supervised training priors.

The application to Advanced LIGO O3 data confirms that the method is effective across the 1-, 3-, and 5-day subset time-spans considered here. The pipeline rescaled all spectral features that it identified and classified as incoherent, corresponding to approximately 89\% and 77\% of the detected features in H1 and L1, respectively, toward the fitted noise baseline. Crucially, statistical verification via Q-Q analysis confirms that this mitigation is achieved without introducing artificial nulls or discontinuities or attenuating potential signals. The retention of coherent spectral features, including the prominent structure near 510~Hz, illustrates how the network-coherence criteria preserve coincident features while targeting detector-specific artefacts. Both the O3 hardware injections and simulated software injections were consistently recovered, with minimal loss in $\cF$-statistic, outside of cleaned frequency bands. Injected signals inside cleaned bands are impacted (in terms of loss in $\cF$-statistic), although in most cases that impact may not wholly prohibit detection. Comparison to the list of O3 line artefacts in~\cite{O3-lines} demonstrates that our automated algorithm identifies a good fraction (but not all) of those lines, consistent with the shorter timescale over which PSDs are averaged ($\le 5$ days) compared to~\cite{O3-lines}.

A key advantage of this framework over purely data-driven machine learning approaches lies in its interpretability and robustness to out-of-distribution artefacts. While deep learning models often struggle with the evolving non-stationarity of the detector state, our method relies on the fundamental physical principle of inter-detector coherence. The statistical consistency of the cleaned background, characterised by a tail slope of $s \approx 0.69$, indicates a conservative rescaling approach that prioritises data continuity over aggressive removal, minimising the risk of false dismissals in the presence of unmodeled noise.

Extensions to the pipeline could incorporate time-dependent coherence tracking to address transient spectral artefacts that evolve on timescales shorter than the observation baseline. The pipeline might also serve as a preprocessing stage for deep-learning-based de-noising algorithms. By removing the dominant stationary lines, our method can reduce the complexity of the feature space, potentially enhancing the generalisation capability of neural networks in extracting sub-threshold astrophysical CW signals.

There remain a few areas where further evaluation of the pipeline would be desirable. We have not tested the pipeline using CW analysis methods other than the $\cF$-statistic. Many CW search methods use coherent integration times shorter than the minimum one-day PSD-averaging time-scale considered in this study. It would therefore be important to characterise the effectiveness of the cleaning pipeline when the PSD used to construct the cleaning mask is averaged over shorter time-scales. CW search methods based on HMM tracking are particularly susceptible to instrumental artefacts; by allowing the CW signal frequency to wander stochastically, the method can easily latch onto loud instrumental disturbances. For such search methods, the line cleaning pipeline we have developed could be particularly helpful.

We envision a comprehensive ``mock data challenge'' to further test the pipeline. First, a population of simulated CW signals would be added to GW data containing line artefacts; input data to CW search methods (e.g. SFTs) would be generated from this uncleaned data. The uncleaned data containing simulated signals would then be processed by the pipeline to generate equivalent cleaned data sets; different choices of the PSD averaging timescale (cf. 1--5 days) could be explored here. Further CW search input data sets would be generated from the cleaned data sets. Finally, a variety of CW search methods would be applied to the uncleaned and cleaned data sets, and the results compared. Comparison metrics of interest would include the fraction of simulated CW signals recovered in the cleaned data compared to the uncleaned data. Ideally, the CW signals recovered in the cleaned data would be a strict super-set of those recovered in the uncleaned data, demonstrating not only that the pipeline does not falsely veto CW signals, but may indeed enable the detection of weak CW signals masked by strong line artefacts that may otherwise have been missed. We defer this proposal to future work.

\ack{We thank Ling Sun for valuable comments and suggestions on the early stages of this project. We thank Evan Goetz for valuable comments on the manuscript. We thank the anonymous referees for constructive comments and suggestions. This work was performed on the OzSTAR national facility at Swinburne University of Technology. The OzSTAR program receives funding in part from the Astronomy National Collaborative Research Infrastructure Strategy (NCRIS) allocation provided by the Australian Government, and from the Victorian Higher Education State Investment Fund (VHESIF) provided by the Victorian Government. This research has made use of data or software obtained from the Gravitational Wave Open Science Center (gwosc.org), a service of the LIGO Scientific Collaboration, the Virgo Collaboration, and KAGRA. This material is based upon work supported by NSF's LIGO Laboratory which is a major facility fully funded by the National Science Foundation, as well as the Science and Technology Facilities Council (STFC) of the United Kingdom, the Max-Planck-Society (MPS), and the State of Niedersachsen/Germany for support of the construction of Advanced LIGO and construction and operation of the GEO600 detector. Additional support for Advanced LIGO was provided by the Australian Research Council. Virgo is funded, through the European Gravitational Observatory (EGO), by the French Centre National de Recherche Scientifique (CNRS), the Italian Istituto Nazionale di Fisica Nucleare (INFN) and the Dutch Nikhef, with contributions by institutions from Belgium, Germany, Greece, Hungary, Ireland, Japan, Monaco, Poland, Portugal, Spain. KAGRA is supported by Ministry of Education, Culture, Sports, Science and Technology (MEXT), Japan Society for the Promotion of Science (JSPS) in Japan; National Research Foundation (NRF) and Ministry of Science and ICT (MSIT) in Korea; Academia Sinica (AS) and National Science and Technology Council (NSTC) in Taiwan. LIGO Document P2600097.}

\funding{This research is supported by the Australian Research Council Centre of Excellence for Gravitational Wave Discovery (OzGrav), project number CE230100016.}

\appendix

\section{Appendix}

Table~\ref{tab:coherent-lines} lists the coherent lines typically found by this work, and in~\cite{O3-lines}.

\begin{table}
  \centering
  \begin{tabular}{rrrr}
    \hline
    \hline
    Min.\ Frequency (Hz) & Max.\ Frequency (Hz) & Bandwidth (Hz) & Source \\
    \hline
    13.40264 & 14.40264 & 1.00000 & O3 lines list \\
    15.09986 & 15.10014 & 0.00028 & O3 lines list \\
    27.47038 & 27.51408 & 0.04370 & O3 lines list \\
    33.19986 & 33.20014 & 0.00028 & O3 lines list \\
    40.86455 & 40.89655 & 0.03200 & O3 lines list \\
    40.89944 & 40.91967 & 0.02023 & O3 lines list \\
    59.99500 & 60.00500 & 0.01000 & This work \\
    99.98500 & 100.01500 & 0.03000 & This work \\
    119.99000 & 120.01000 & 0.02000 & This work \\
    179.99500 & 180.00500 & 0.01000 & This work \\
    314.76153 & 315.32153 & 0.56000 & O3 lines list \\
    504.25945 & 504.50568 & 0.24623 & O3 lines list \\
    504.91500 & 511.68500 & 6.77000 & This work \\
    507.89972 & 516.25972 & 8.36000 & O3 lines list \\
    819.50000 & 832.30000 & 12.80000 & This work \\
    910.15000 & 918.85000 & 8.70000 & This work \\
    999.61265 & 999.64065 & 0.02800 & O3 lines list \\
    1006.38262 & 1006.49962 & 0.11700 & O3 lines list \\
    1006.69044 & 1006.77005 & 0.07961 & O3 lines list \\
    1010.97520 & 1011.17220 & 0.19700 & O3 lines list \\
    1011.30193 & 1011.36893 & 0.06700 & O3 lines list \\
    1014.81149 & 1014.83549 & 0.02400 & O3 lines list \\
    1016.17786 & 1016.23986 & 0.06200 & O3 lines list \\
    1016.34055 & 1016.61755 & 0.27700 & O3 lines list \\
    1017.89965 & 1017.91565 & 0.01600 & O3 lines list \\
    1083.10000 & 1083.70000 & 0.60000 & This work \\
    1153.09986 & 1153.10014 & 0.00028 & O3 lines list \\
    1471.00605 & 1471.40605 & 0.40000 & O3 lines list \\
    1482.53359 & 1482.58359 & 0.05000 & O3 lines list \\
    1494.26147 & 1494.37263 & 0.11116 & O3 lines list \\
    1499.55471 & 1499.62199 & 0.06728 & O3 lines list \\
    1666.71000 & 1667.29000 & 0.58000 & This work \\
    1825.10000 & 1858.10000 & 33.00000 & This work \\
    1932.38000 & 1961.62000 & 29.24000 & This work \\
    1936.15625 & 1936.36403 & 0.20778 & O3 lines list \\
    \hline
    \hline
  \end{tabular}
  \caption{Coherent lines typically found by this work, and in the O3 lines list.}
  \label{tab:coherent-lines}
\end{table}

\providecommand{\newblock}{}

\end{document}